\documentclass[a4paper,showpacs,twocolumn,amsmath,amssymb,floatfix,prb,preprintnumbers,footinbib]{revtex4}
\usepackage{graphicx}
\usepackage{amsmath,amsfonts}

\newcommand{\e}{\varepsilon}
\newcommand{\m}[1]{\mathrm{#1}}
\newcommand{\bo}[1]{\boldsymbol{#1}}
\newcommand{\ket}[1]{|#1\rangle}
\newcommand{\bra}[1]{\langle #1|}

\begin{document}

\title{Interference and interaction effects in adiabatic pumping through quantum dots}

\author{Bastian Hiltscher$^1$, Michele Governale$^2$, and J\"urgen K\"onig$^1$} 
\affiliation{$^1$Theoretische Physik, Universit\"at Duisburg-Essen and CeNIDE, 47048 Duisburg, Germany\\
$^2$School of Chemical and Physical Sciences  and MacDiarmid Institute for Advanced Materials and Nanotechnology, Victoria University of Wellington, P.O. Box 600, Wellington 6140, New Zealand
}
 
\date{\today}
%abstract
\begin{abstract}
In order to investigate the effects of interference and interaction in adiabatic pumping, 
we consider an Aharonov-Bohm (AB) interferometer with a quantum dot embedded either in one or in both arms. We employ a real-time formalism and we perform an expansion both in the tunnel-coupling strengths between dot and leads and in the pumping frequency, taking into account the Coulomb interaction non perturbatively. We find that pumping in a single-dot AB interferometer has a peristaltic but phase-coherent character. 
In a double-dot AB interferometer, we find a pumping mechanism that relies purely on 
quantum-mechanical interference and has no classical counterpart.

\end{abstract}
% 73.23.Hk 	Coulomb blockade; single-electron tunneling 
%85.35.Ds 	Quantum interference devices 
%72.10.Bg 	General formulation of transport theory 

\maketitle
%introduction
\section{Introduction}
Multiply connected mesoscopic structures are ideal to study quantum interference in a solid-state system. In an Aharonov-Bohm (AB) geometry, the sensitivity of the current to the magnetic flux enclosed by the two interfering paths can be used as a measure of the quantum coherence in the system. Embedding quantum dots in the arms of the  interferometer allows the investigation of the coherence of transport through a region with strong Coulomb interaction.  Several experiments\cite{yacoby94,schuster97,ji00,wiel00,holleitner01} have confirmed that the visibility of AB oscillations in the current is not completely suppressed by the presence of a quantum dot, indicating that transport though the strongly-interacting dot is partially coherent.  
One mechanism of decoherence that has been investigated both experimentally \cite{aikawa04} and theoretically,\cite{koenig} is spin-flip tunneling. 

The goal of the present paper is to address the issue of coherence in adiabatic pumping through systems with strong Coulomb interaction. To this end, we consider pumping in AB-interferometer devices with a quantum dot embedded either in one or in both arms.

Pumping is a transport mechanism, which exploits the periodic time dependence of some parameters of a nanoscale conductor to produce a dc current in the absence of an applied bias voltage. 
Its appeal for both theorists and experimentalists lies in the possibilities it offers to investigate the non-equilibrium induced by the explicit time-dependence of a nanoscale system. The adiabatic-pumping regime is  characterized by the pumping frequency being smaller than the characteristic time scales of the system. Recently, there have been  several experiments on pumping in nano systems.  \cite{pothier92,martinis,switkes99,watson,fletcher,fuhrer,buitelaar,kaestner}
A lot of the theoretical effort has been devoted to systems, where the Coulomb interaction can be treated within a mean-field approach.\cite{brouwer98,zhou99,buttiker01,makhlin01,buttiker02,entin02,moskalets02,luis02,arrachea06,moskalets08} In this regime, a well established theoretical framework for pumping, based on the dynamical scattering approach to mesoscopic transport,  exists.\cite{brouwer98,buettiker94}
However, in some nano-scale systems, such as few-electron quantum dots, Coulomb interaction can become very important, requiring a non-perturbative treatment. In the last few years, pumping in strongly interacting systems has attracted a lot of theoretical interest. \cite{aleiner98,citro03,aono04,cota05,brouwer05,splett05,sela06, splett06,sanchez,fioretto07,braun08,arrachea08,splett08, cavaliere, hernandez09}
In the present paper, we employ a diagrammatic approach to adiabatic pumping in quantum 
dots,\cite{splett06} which relies on a systematic expansion in both the pumping frequency and the tunnel-coupling strengths. This formalism is valid in the weak-tunneling regime but it takes into account the on-site Coulomb repulsion in the dot non-perturbatively. 

Sometimes the term {\it adiabatic quantum pumping} is used in the literature to emphasize 
the role of quantum interference in a given pumping mechanism, in contrast to a purely classical 
pump. This motivates the question whether such a distinction is well defined. We find that this is not always the case. For, example, the pumping mechanism we discuss in the case of a single-dot AB interferometer exhibits both quantum and classical features at the same time.
For the case of a double-dot AB interferometer, on the other hand, we identify a pumping scheme that relies exclusively on quantum-mechanical interference.

An important issue when interpreting experimental data is to distinguish pumping from 
rectification.\cite{brouwer01} In fact, due to the presence of stray capacitances, undesired ac bias voltages may appear across the time-dependent conductor, and can give rise to a dc current. 

In this paper, we compare pumping and rectification, analyze the different processes contributing to transport and discuss to which degree symmetry with respect to the magnetic field can be used to distinguish the two transport mechanisms.

The paper is organized as follows. In Sec.~\ref{model}, we present the model and the technique used to compute the pumping current. The results for an interferometer with one quantum dot embedded in only one of the arms and with a quantum dot in both arms are presented in Secs.~\ref{results1dot} and \ref{results2dots}, respectively. Finally, conclusions are drawn in Sec.~\ref{Conclusions}.

\begin{figure}[h]
\centering
\includegraphics[width=7cm]{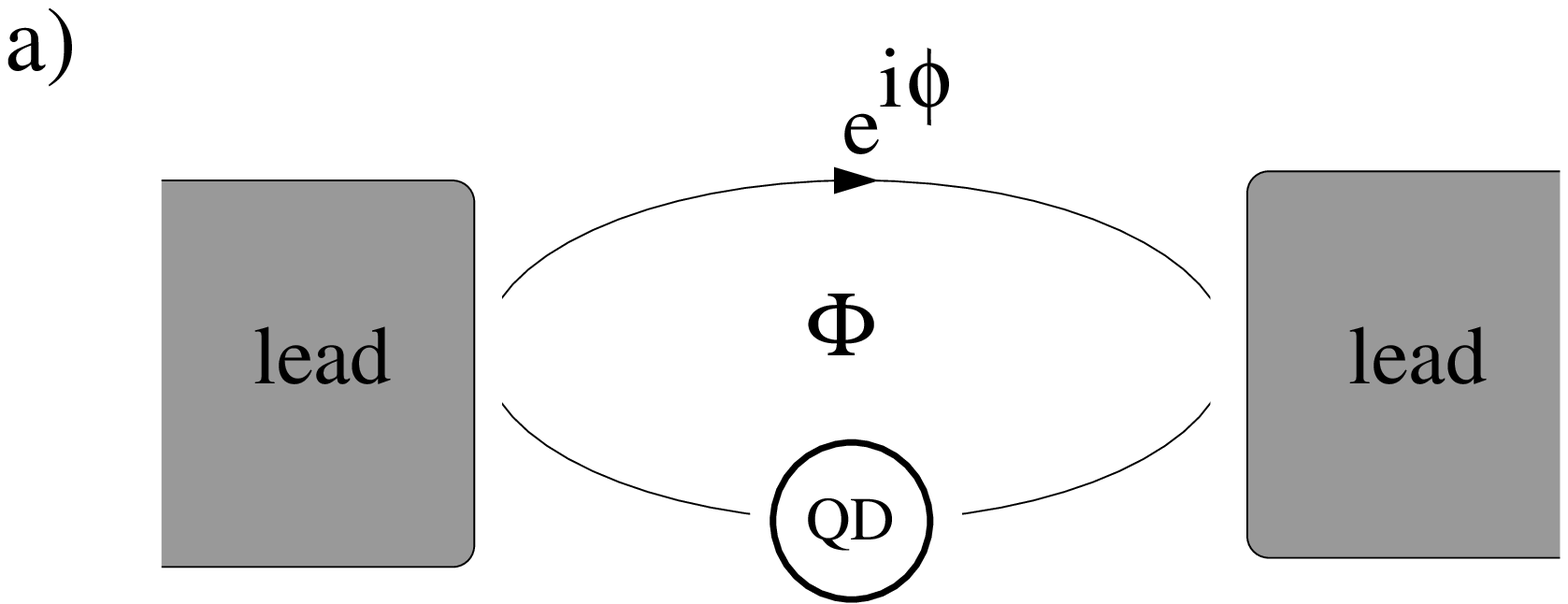}
\includegraphics[width=7cm]{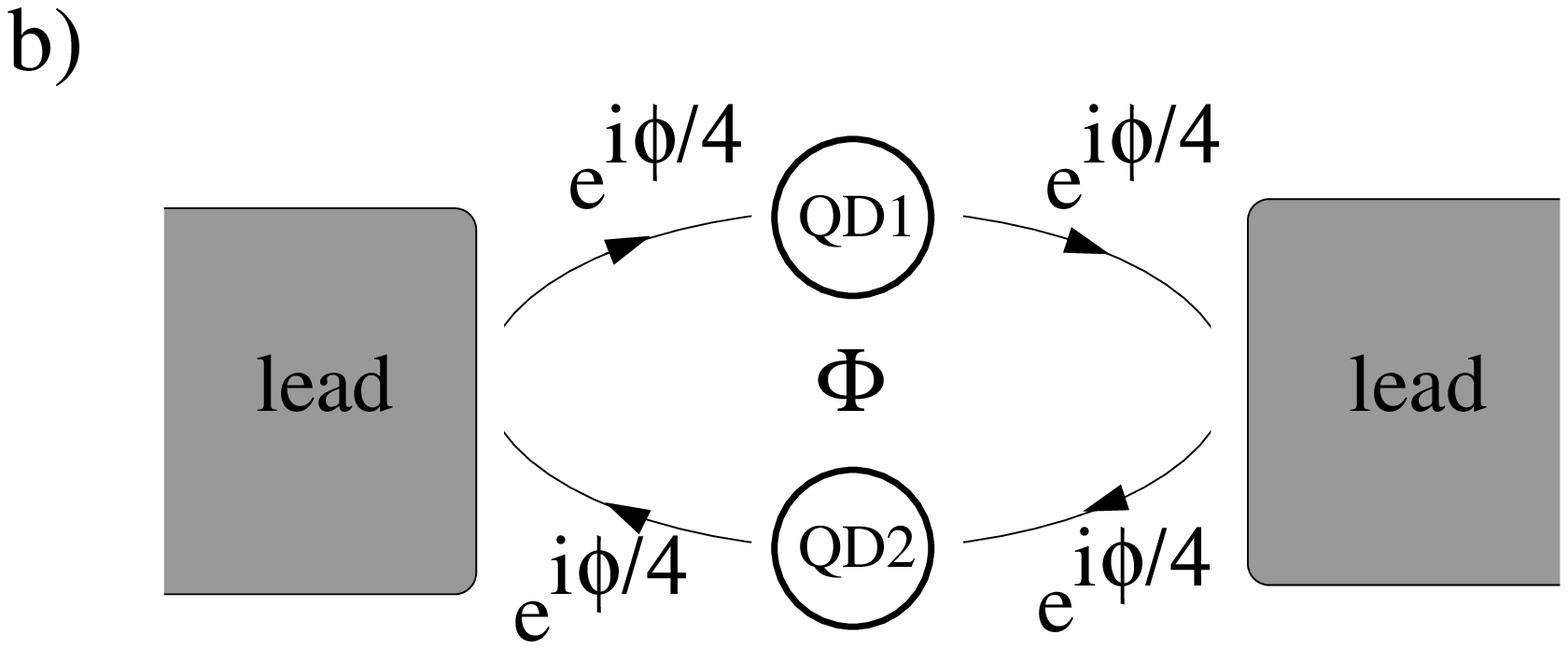}
\caption{\label{setup}Setup of (a) single-dot and (b) double-dot Aharonov-Bohm interferometer.
}
\end{figure}

%theory
%hamiltonian
\section{Model and Theoretical approach}
\label{model}
\subsection{Model}\label{sec_hamiltonian}
We start by defining the different building blocks of the quantum-dot Aharonov-Bohm interferometers depicted in Fig.~\ref{setup}.
The quantum dots, numbered by the index $j$, are assumed to be in the single-level regime, i.e., they can be viewed as Anderson impurities, 
\begin{equation}
\label{dothamil}Ê 
H_{\text{dot},j}=\e_j\sum\limits_\sigma n_{j\sigma} + Un_{j\uparrow} n_{j\downarrow \,} .
\end{equation}
Here, $n_{{j\sigma}}=d_{j\sigma}^\dagger d_{j\sigma}$, with $d_{j\sigma}^\dagger $ being the creation operator for an electron with spin $\sigma$ in quantum dot $j$.  The dot-level position is denoted by $\epsilon_j$ and the onsite Coulomb-repulsion energy by $U$.  The inter-dot charging energy for a double-dot interferometer is assumed to be negligible.
The two leads are modeled as reservoirs of non-interacting electrons,
\begin{equation}
H_{\text{leads}}=\sum\limits_{r k\sigma}\e_{rk}c_{rk\sigma }^\dagger c_{rk\sigma
}\, ,
\end{equation}
where  $ c_{r k\sigma}^\dagger$ is the creation operator for an electron in lead $r=\text{L, R}$ in a state labeled by the quantum number $k$ and with spin $\sigma$. 

The tunnel coupling between dot $j$ and the leads is modeled by the tunneling Hamiltonian,
\begin{equation}
\label{tunnelhamil}
H_{\rm{tunn},j}=\sum\limits_{k\sigma r}t_{rj}\left(  c_{r k\sigma}^\dagger d_{j\sigma}
+\mathrm{H.c.} \,\right) .
\end{equation}
We  assume the tunnel matrix elements $t_{rj}$ and the density of states $N_r$ in the lead $r$ to be energy independent in the energy window relevant for transport. Tunnel-coupling strengths are then defined as $\Gamma_{r j}=2\pi |t_{rj}|^2 N_r$.
Furthermore, we define $\Gamma_{j}=\sum_r \Gamma_{r j}$.

An interferometer arm without a quantum dot (reference arm) is modeled by a direct tunnel coupling between the leads,
 \begin{eqnarray}
H_{\text{ref}}=\sum\limits_{k \in R, q \in L, \sigma} (\tilde{t}c^\dagger_{\text{R} k\sigma}c_{\text{L} q\sigma }+\m{H.c.})\, ,
\end{eqnarray}
with transmission amplitude $t^{\text{ref}}=2\pi \sqrt{N_{\text{L}}N_{\text{R}}}\, \tilde{t}$.

In this paper we focus on two different setups:

(i)  A single-dot Aharonov-Bohm interferometer, as shown in Fig.\ref{setup} (a). In this case, in order to avoid cluttering the notation, we consistently drop the dot index $(j=1)$. 
The Hamiltonian is simply $H=H_{\text{dot}}+H_{\text{tunn}}+H_{\text{leads}}+ H_{\text{ref}}$. The magnetic flux $\Phi$  threading the interferometer is included in the phases of the tunneling amplitudes. We choose the gauge in which $t_{\text{L}},t_{\text{R}} \in \Re^+$ and $\arg \tilde{t}=\phi=2\pi \Phi/\Phi_0$, where $\Phi$ is the magnetic flux and $\Phi_0$ is the flux quantum.

(ii)  A double-dot Aharonov-Bohm interferometer, as shown in Fig.\ref{setup} (b). 
In this case, the Hamiltonian reads 
 $H=\sum_{j=1,2} \left[H_{\text{dot},j} +H_{\text{tunn},j}\right] +H_{\text{leads}}$. 
 We choose the gauge in which 
 $-\arg t_{\text{L}1}=\arg t_{\text{R}1}=-\arg t_{\text{R}2}=\arg t_{\text{R}1}=\phi/4$.

%real-time perturbation theory
\subsection{Real-time diagrammatic approach to pumping}
We generalize the real-time diagrammatic approach to pumping introduced in Ref.~\onlinecite{splett06} for a single-dot or double-dot Aharonov-Bohm interferometer.
New ingredients are the possibility of direct tunneling from source to drain for the reference arm in the single-dot case and the inclusion of off-diagonal elements of the reduced density matrix which account for coherent superpositions of the electrons in different quantum dots in the double-dot case.

The systems under consideration can be decomposed in a subsystem with few degrees of freedom comprising  the quantum dots, and  the leads which possess a large number  of non-interacting   degrees of freedom.  
Since we are not interested in the dynamics of the leads' degrees of freedom, we can trace them out thus obtaining an effective description of the quantum dots' subsystem. 
The Hilbert space of the reduced system is spanned by the eigenstates $\ket{\chi}$  of the Hamiltonian of the isolated dot(s), $\sum_j H_{\text{dot},j}$. The corresponding eigenenergies are denoted by $E_\chi$.
For the single-dot case, a natural choice for the basis $\{\ket{\chi}\}$ is $\{\ket{0},\ket{\uparrow},\ket{\downarrow},\ket{\uparrow \downarrow}\equiv d_\uparrow^\dagger d_\downarrow^\dagger \ket{0}$\}, where the different states correspond, respectively, to empty, singly occupied by spin up, singly occupied by spin down, and doubly occupied dot. The dynamics of the dots' degrees of freedom are fully described by the reduced density matrix $\bf{p}$, whose matrix elements are $p_{\chi_2}^{\chi_1}=\bra{\chi_1}\mathbf{p}\ket{\chi_2}$. A diagonal element of the reduced density matrix $p_\chi^\chi$ denotes the probability of the dot being in a state $\chi$.
We introduce the vector $\boldsymbol{\pi}=(p_{\chi_1}^{\chi_1},...,p_{\chi_m}^{\chi_m},...,p_{\chi_j}^{\chi_i},...)^\m{T}$, (with $i\neq j$), whose first $m$ components are all diagonal elements of the reduced density matrix followed by all off-diagonal elements. 
The dynamics of reduced system is governed by the generalized master equation (in matrix notation)
\begin{equation}\label{genmastereq}
	\frac{d}{d t}\bo{\pi}(t)=-\m{i}{\bf E}(t)\bo{\pi}(t)+\int\limits^t_{-\infty}d t'{\bf W}(t,t')\bo{\pi}(t')\, ,
\end{equation}
where the Kernel element $W_{\chi'\chi'''}^{\chi\,\,\,\chi''}(t,t')$ describes the transition  from an initial state described by $p_{\chi'''}^{\chi''}$ to a final state described by $p_{\chi'}^{\chi}$. The matrix elements of ${\bf E}(t)$ are given by $E_{\chi'\chi'''}^{\chi\chi''}(t)=\delta_{\chi\chi''}\delta_{\chi'\chi'''}\left(E_\chi(t)-E_{\chi'}(t)\right)$. \cite{note}

We are interested in pumping, i.e., in transport due to the periodic variation of the system parameters, collectively denoted by $X$. The vector $\bo{\pi}(t)$ as well as the kernel ${\bf W}(t,t')$ depend in a functional way on the pumping parameters $X(\tau)$. To solve Eq. (\ref{genmastereq}) we perform an adiabatic expansion, \cite{splett06}  i.e., an expansion in powers of the pumping frequency $\Omega$, which is 
valid when the pumping frequency is much smaller than the response time of the system. For this purpose, we first perform a Taylor expansion around the final time $t$ of $\bo{\pi}(\tau)=\bo{\pi} (t)+(\tau-t)\frac{d\bo{\pi}}{d\tau}(\tau)\Big|_{\tau=t}$ in the integral on the right-hand side of  Eq. (\ref{genmastereq}). 
Furthermore, we need to perform the adiabatic expansion of the kernel  ${\bf W}(t,t')$. In order to do so, 
we expand the parameters around the time $t$, i.e., $X(\tau)=X(t)+(\tau-t)\frac{d}{d \tau}X(\tau)\Big|_{\tau=t}$.
We write the kernel expansion as 
\begin{equation}
	{\bf W}(t,t')={\bf W}^\m{(i)}_t(t-t')+{\bf W}^\m{(a)}_t(t-t').
\end{equation}
The subscript $t$ denotes the time $t$ around which the adiabatic expansion is performed.
The instantaneous part [with superscript $(i)$] is obtained by freezing all parameters at time $t$. The adiabatic correction term [with superscript $(a)$] contains only terms which are linear in time derivatives of the pumping parameters  $\frac{d}{d \tau}X(\tau)\Big|_{\tau=t}$. 
Finally, we perform an adiabatic expansion of the reduced density matrix,
\begin{eqnarray}
	\bo{\pi}(t)=\bo{\pi}^\m{(i)}_t+\bo \pi^\m{(a)}_t \, .
\end{eqnarray}
The instantaneous part can be obtained by solving the generalized master equation in the stationary limit
\begin{eqnarray}\label{instgenmastereqn}
	0=\left( -\m{i}{\bf E}(t)+{\bf W}^\m{(i)}_t \right) \bo{\pi}^\m{(i)}_t
\end{eqnarray}
together with the normalization condition $\bo n \bo \pi_t^\m{(i)}=1$ with $\bo n=(1,...1,0,...,0)$, i.e., the first $m$ components of $\bo n$ are $1$ and the other components are $0$. In Eq.~(\ref{instgenmastereqn}), we have introduced the generalized rates as the Laplace transform of the Kernel computed at zero frequency:
${\bf W}^\m{(i/a)}_t=\lim\limits_{z\rightarrow0^+} \int \limits_{-\infty}^t d t' e^{-z(t-t')}{\bf W}^\m{(i/a)}_t(t-t')$.

The first adiabatic correction of the generalized master Eq. (\ref{genmastereq}) reads
\begin{equation}\label{adiabgenmastereqn}
	\frac{d}{dt}\bo\pi_t^\m{(i)}=\left(-\m{i}\bf E(t)+{\bf W}^\m{(i)}_t\right)\bo \pi^\m{(a)}_t+{\bf W}^	\m{(a)}_t\bo \pi_t^\m{(i)} +\partial {\bf W}^{(i)}_t\frac{d}{dt}\bo \pi_t^\m{(i)}\, ,
\end{equation}
with $\partial{\bf W}^{(\m{i})}_t=\lim \limits_{z\rightarrow0^+} \frac{d}{d z}\int \limits_{-\infty}^t d t' e^{-z(t-t')}{\bf W}^{(\m{i})}_t(t-t')$. Equation (\ref{adiabgenmastereqn}) together with the normalization condition   $\bo n \bo \pi_t^\m{(a)}=0$ allows to determine the adiabatic correction of the reduced density matrix $\bo \pi^{(a)}_t$.

In the following, we concentrate on the limit of weak tunnel couplings.
Therefore, we perform a perturbation expansion in the tunnel-coupling strength $\Gamma$ between dot and leads. The $k$th order contribution to the reduced density matrix is denoted by $\bo \pi^{(\m{i/a},k)}_t$. 
Matching the orders in Eq.~(\ref{adiabgenmastereqn}), it is easy to see that the expansion of the instantaneous term of the reduced density matrix $\bo\pi^{(\m{i},k)}_t$ starts in zeroth order ($k=0$), while the adiabatic correction $\bo\pi^{(\m{a},k)}_t$ starts in minus first order in $\Gamma$ ($k=-1$).
This does not invalidate the expansion, since due to the low-frequency condition $\Omega\ll \Gamma$ the correction $\bo\pi^{(\m{a},-1)}_t\propto \Omega/\Gamma$ still remains small. 
In the single-dot AB interferometer, we also need to consider direct tunneling processes between the two leads. This is done by performing a perturbation expansion in $|t^\m{ref}|$. The order of  $|t^\m{ref}|$ is indicated by another superscript $l$, i.e., the corrections to the reduced density matrix are now  denoted by $\bo\pi^{(\m{i/a},k,l)}_t$.

The expectation value of the current flowing into lead $r$ consists of an instantaneous part and its adiabatic correction. The instantaneous part reads
\begin{eqnarray} 
	J_r^{(\m{i})}(t)=e{\bo n}{\bf W}^{r,\m{(i)}}_t\bo\pi_t^{\m{(i)}}\, ,
\end{eqnarray}
with the current rates ${\bf W}^{r}$.
The latter are calculated similar to ${\bf W}$ but are weighted with the number of electrons transferred to lead $r$. 
Without applied bias voltage the instantaneous part  vanishes. 
Hence, the pumped current is given by the adiabatic correction 
\begin{equation}
\label{adiabcurr}
	J_r^\m{(a)}(t)=e{\bo n}\left({\bf W}^{r,\m{(a)}}_t\bo \pi_t^\m{(i)}+{\bf W}^{r,\m{(i)}}_t\bo \pi_t^	\m{(a)}+\partial{\bf W}^{r,\m{(i)}}_t\frac{d}{dt}\bo\pi_t^\m{(i)}\right)\, ,
\end{equation}
where  the superscript ${r}$ points out that the rates, both the instantaneous ones and their adiabatic corrections, are  current rates and the number of transferred electrons needs to be accounted for.

%results
\section{Results}
Using the real-time perturbation theory outlined in the previous section, we compute the pumped current through an AB interferometer with a quantum dot embedded either in one or in both arms.
We concentrate on the limit of weak tunnel coupling and we expand the pumped current up to the lowest order that is AB flux dependent, which is associated with AB interference.
Furthermore, we calculate the dc current driven through the AB interferometer by an applied ac bias voltage and rectified by the time dependence of the instantaneous conductance of the system. We identify the characteristic features for both transport mechanisms. This  helps us to deepen our understanding of pumping, but more importantly, to distinguish pumping from rectification in an experiment.
%single-dot ab interferometer
%pumped current
\subsection{Single-dot AB interferometer}
\label{results1dot}
\subsubsection{Weak adiabatic pumping}
Without applied bias voltage, the instantaneous part of the current vanishes. The lowest order of the adiabatic correction,
\begin{align}
\label{1dotcurrloworder}
	J_\m{L}^\m{(a,0,0)}(t)= -{e}\frac{\Gamma_\m{L}}{\Gamma}\frac{d}{d t} \left
	\langle n \right \rangle^\m{(i,0,0)}\, ,
\end{align}
is proportional to the time derivative of the average dot occupation $\langle n\rangle^{(i,0,0)} =\frac{2f(\e)}{1+f(\e)-f(\e+U)}$, where $f(\e)$ is the Fermi distribution.
Equation (\ref{1dotcurrloworder}) has very simple interpretation:
if the average occupation changes as a result of varying  the dot level $\e$, then charge flows into or out of the dot. The fraction of the current flowing through the left lead is simply given by the ratio
\begin{align}
\label{ratio1}
\frac{W_{\chi'\leftarrow\chi}^\m{L,(i,1,0)}}{W^\m{(i,1,0)}_{\chi'\leftarrow\chi}}=\frac{\Gamma_\m{L}}{\Gamma}
\end{align}
where $W_{\chi'\leftarrow\chi}^\m{L,(i,1,0)}$ is the golden-rule rate  for a transition from $\chi$ to $\chi'$ with an electron tunneling through the left barrier, while  $W^\m{(i,1,0)}_{\chi'\leftarrow\chi}$ is the total rate for tunneling through the left and right barriers. 
Notice that in the lowest-order perturbation theory considered here, the golden-rule rates coincide with the generalized rates introduced in the previous section. 
The ratio Eq.~(\ref{ratio1}) is independent from the initial and final states, $\chi$ and $\chi'$. 

The first flux-dependent correction to the current is 
\begin{align}
	\label{korrekturstrom1dot} 
	J_\m{L}^\m{(a,0,1)}(t)={e} \frac{\sqrt{\Gamma_\m{L}
	\Gamma_\m{R}}|t^{\mathrm{ref}}|}{\Gamma}\sin\phi \frac{d}{d t}\left
	\langle n \right \rangle^\m{(i,0,0)}\, .
\end{align}
It can be interpreted in a similar way as the lowest-order contribution with the only difference that now the ratio between the tunneling rates is given by
\begin{align*}
	\frac{W_{\chi'\leftarrow\chi}^\m{L,(i,1,1)}}{W^\m{(i,1,0)}_{\chi'\leftarrow\chi}}=-\frac{\sqrt{\Gamma_\m{L}
	\Gamma_\m{R}}|t^{\mathrm{ref}}|}{\Gamma}\sin\phi\, .
\end{align*} 
Those rates describe the flux-dependent parts of the processes that fill or empty the dot.
The flux dependence arises due to the interference of the two possible paths available: either direct tunneling between dot and lead $r$ or the indirect paths that transfers an electron between dot and lead $r$ via the other lead $\bar r$.
These rates, that change the dot occupation, have quite different properties than the rates of same order that describe transfer of electrons from one lead to the other.\cite{urban08} 
In particular, they are odd functions of the AB flux. 

In the adiabatic regime, to obtain a non-vanishing pumped charge at least two parameters ($X_1$ and $X_2$) need to be time dependent.  We write the parameters as $X_i (t)=\bar{X}_i+\delta X_i(t)$, where $\bar{X}_i$ is the mean value and $\delta X_i(t)$ is the oscillating component. 
We indicate the pumped charge due to the variation of $X_1$ and $X_2$ as 
$Q_{X_1,X_2}$;  it can be compute as $ Q_{X_1,X_2}=\int\limits_0^{2\pi/\Omega} d tJ_\m{L}(t)$.
In the following, we consider weak pumping and compute the pumped charge in bilinear order in  
$\delta X_i(t)$.
Since the current is proportional to the time derivative of the average dot occupation, see Eqs. (\ref{1dotcurrloworder}) and (\ref{korrekturstrom1dot}), one pumping parameter needs to be the dot level $\e$.  Choosing $\Gamma_\m{L}$ to be the second pumping parameter we obtain for the pumped charge in zeroth order in $\Gamma$ and in zeroth and first order in $|t^{\text{ref}}|$ 
\begin{align}
	\label{gammaL} 
		Q_{\Gamma_\m{L},\e}^\m{(a,0,0)} &=
	-{e}\frac{\Gamma_\m{R}}{\bar{\Gamma}^2} \eta_{\Gamma_\m{L},\e}
	\frac{d}{d\bar{\e}} \left \langle \bar{n} \right \rangle^\m{(i,0,0)}\, ,\\
	\label{gammaL1}
	Q_{\Gamma_\m{L},\e}^\m{(a,0,1)}&=
	{e}\sqrt{\frac{\Gamma_\m{R}}{\bar{\Gamma}_\m{L}}}\frac{\Gamma_\m{R}-\bar{\Gamma}_\m{L}}	{2\bar{\Gamma}^2}|t^{\mathrm{ref}}|\eta_{\Gamma_\m{L},\e}
	\sin \phi  \frac{d}{d\bar{\e}} \left \langle \bar{n}
	\right \rangle^\m{(i,0,0)}\, ,
\end{align}
where the prefactor
\begin{align*}
	\eta_{\Gamma_r,\e}=\int \limits_0^{2\pi/\Omega} \frac{\partial
	\delta\e}{\partial t}\delta \Gamma_r d t
\end{align*}
is the area of the pumping cycle in parameter space. The charge for pumping with $(\Gamma_{\text{R}},\e)$ is simply obtained from Eq.~(\ref{gammaL}) swapping $\Gamma_{\text{L}}$ and $\Gamma_{\text{R}}$.

\begin{figure}[h]
\centering
\includegraphics[width=7cm]{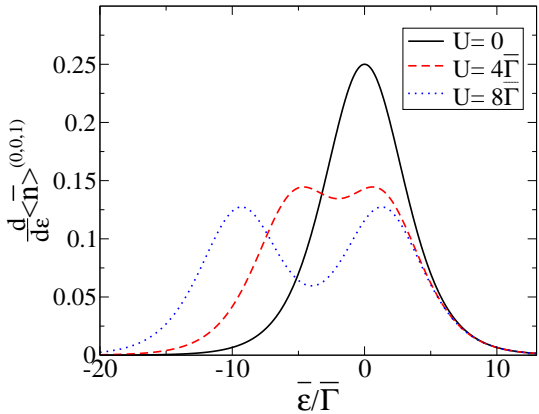}
\caption{\label{eindotpumpen} 
The dependence of pumped charge in zeroth order in $\Gamma$ and up to first oder in $|t^\m{ref}|$ on the average dot level $\bar{\e}$  is given by the derivative of the average dot occupation $\frac{d}{d\bar{\e}} \left \langle \bar{n}
\right \rangle^\m{(i,0,0)}$, which is plotted here for various Coulomb interaction strengths $U$. The temperature is $k_\m{B}T=2\bar{\Gamma}$.
} 
\end{figure}

The dependence of the pumped charge on the average level position in zeroth order in $\Gamma$ and up to first order in $|t^{\text{ref}}|$, see Eqs. (\ref{gammaL}) and (\ref{gammaL1}), is simply given by  $ \frac{d}{d\bar{\e}} \left \langle \bar{n} \right \rangle^\m{(i,0,0)}$.
The latter is plotted in Fig.~ \ref{eindotpumpen}.
The pumped charge is even around $\bar{\e}=-U/2$. 
The direction of the pumped current is independent of the average dot level.
The two peaks in Fig.~ \ref{eindotpumpen} are associated to the transitions between singly occupied and empty dot and between doubly occupied and singly occupied dot.

Is the mechanism of pumping in our example of classical or of quantum nature? 
The interpretation of Eqs.~(\ref{1dotcurrloworder}) and (\ref{korrekturstrom1dot}) is consistent with a picture of a peristaltic pump: a variation of the gate voltage pushes the electrons off the dot, which generates a current flow to the leads, where the larger part flows through the contact that is more open, as characterized by the relative coupling strengths to the left and right leads. If this ratio is changed when the gate voltage sucks electrons into the dot later, then a net charge has been transferred from one lead to the other. This mechanism does not rely on the  formation of a superposition of quantum states.
We refer to it as peristaltic, which is usually associated with classical pumping. 
On the other hand, the underlying tunnel processes that transfer the dot electrons from or to the leads are phase coherent, as signaled by the dependence on the AB phase. They are a consequence of quantum-mechanical interference of the two possible paths between dot and a given lead. 
Therefore, we conclude that a clear distinction between quantum and classical pumping is meaningless  for our example.
To emphasize the coexistence of both classical and quantum features,   
we describe the mechanism considered here as {\it phase-coherent peristaltic pumping}.

\subsubsection{Comparison with rectification}

Adiabatic pumping may be obscured by rectification. 
A time-dependent gate voltage may not only change the level position in the quantum dot but, due to a parasitic capacitive coupling to the leads, give rise to an effective (in-phase) ac bias voltage.
This ac bias voltage can, in turn, yield a dc current component due to the time dependence of the dot-level position.  
In Ref.~\onlinecite{brouwer01}, symmetry with respect to magnetic field has been used to discriminate pumping from rectification.
Assuming that the lever arms between gates and reservoirs are small, one can neglect rectification contributions quadratic in $V(t)$ but to zeroth order in time variation of the system parameters of the pumping region ($\delta \epsilon$ in our case). This is because the effect of the gate-voltage modulation on the dot-level position dominates over the ac voltage due to the parasitic stray capacitance. 
In this limit, the charge transferred in one period by rectification can be computed as 
$Q_ {\text{rec},X}=\int_0^{2\pi/\Omega} G_{\text{L}}(t) V(t) dt$, where $G_{\text{L}}(t)$ is the instantaneous linear conductance and $V(t)$ is the undesired oscillating bias voltage. 
Due to Onsager relations the linear conductance, and, therefore, also the rectification contribution to the transferred charge, is an even function of the magnetic field.
This reasoning, however, is no longer valid when contributions to the rectified current that are non-linear in the parasitic ac bias voltage have to be taken into account.
In fact, magnetic field symmetries for different transport regimes have been extensively investigated experimentally\cite{dicarlo03,leturcq06,angers07} as well as theoretically.\cite{moskalets05,polianski07}
In nonlinear response it has been measured that Coulomb interaction may yield an odd part also in rectification.\cite{angers07} The ratio between odd and even parts strongly depends on the bias mode and the frequency. Especially, in the adiabatic regime the odd part is in general not negligible.\cite{polianski07} 

\begin{figure}[h]
\centering
\includegraphics[width=7cm]{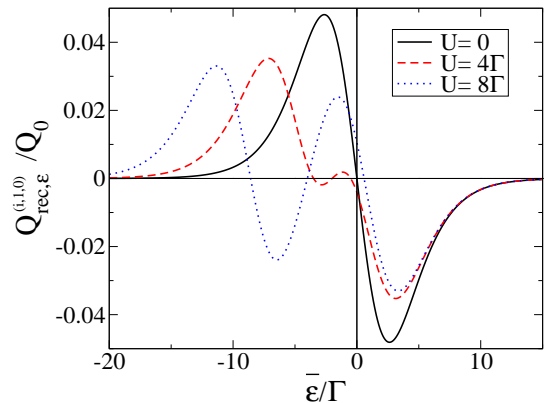}
\caption{\label{recloworder} Charge transferred  by rectification in lowest non-vanishing order ,  $Q_\m{rec,\e}^\m{(i,1,0)}$,  in units of $Q_0={e}^2\frac{\Gamma_\m{R}\Gamma_\m{L}}{\Gamma^3}\eta_{\mathrm{rec},\varepsilon}$ as a function of the average dot level $\bar{\e}$  for various Coulomb-interaction strengths $U$. 
The time-varying parameter is $\e$ and the temperature is $k_\m{B}T=2\bar{\Gamma}$.
}
\end{figure}

In the following, we choose $\e$ as time-dependent parameter and we compute the rectified charge in linear order in $\delta\e(t)$ and $V(t)$.
In lowest non-vanishing order in $\Gamma$,  the charge transferred by rectification reads
\begin{equation}
\begin{split}
\label{qrect}
Q_{\m{rec},\varepsilon}^{(i,1,0)} &=
-e^2\frac{\Gamma_\m{L}\Gamma_\m{R}}{\Gamma}\eta_{\mathrm{rec},\varepsilon}\\
\frac{d}{d\bar{\varepsilon}}&\left[\frac{\left(1-f(\bar{\varepsilon}+U)\right)\frac{d}{d\bar{\varepsilon}}f(\bar{\varepsilon})+f(\bar{\varepsilon})\frac{d}{d\bar{\varepsilon}}f(\bar{\varepsilon}+U)}{1+f(\bar{\varepsilon})-f(\bar{\varepsilon}+U)}\right]
\end{split}
\end{equation}
with $\eta_{\mathrm{rec},\varepsilon}=\int_0^{2\pi/\Omega}d t\delta\e(t)V(t)$. 
The rectified charge Eq.~(\ref{qrect})  is odd around $\bar{\e}=-U/2$ (see Fig \ref{recloworder}).

\begin{figure}[h]
\centering
\includegraphics[width=7cm]{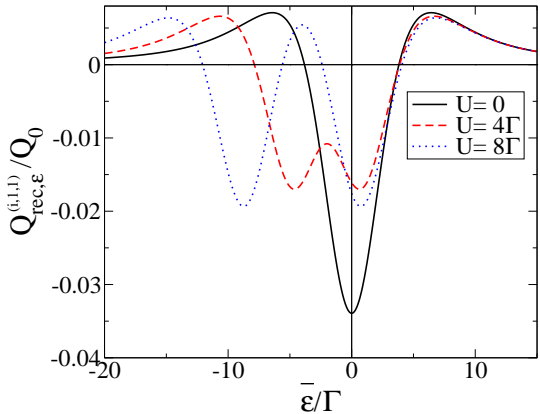}
\caption{\label{reccor} First flux-dependent correction to the charge transferred by  rectification, $Q_\m{rec,\e}^\m{(i,1,1)}$, in units of $Q_0={e}^2\frac{\sqrt{\Gamma_\m{L}\Gamma_\m{R}}}{\Gamma^2}|t^\m{ref}|\eta_{\mathrm{rec},\varepsilon}\cos{\phi}$ as a function of the average dot level $\bar{\e}$ for various Coulomb-interactions strengths $U$. The time-varying parameter is $\e$ and the temperature is $k_\m{B}T=2\bar{\Gamma}$.
} 
\end{figure}

The first flux-dependent correction reads 
\begin{equation}
Q_{\text{rec},\e}^{(i,1,1)} =
-2e^2\frac{\sqrt{\Gamma_\m{L}\Gamma_\m{R}}}{\Gamma}|t^{\mathrm{ref}}|\eta_{\mathrm{rec},\e}
\cos\phi\frac{d}{d\bar{\e}} \left \langle \bar{n} \right\rangle^{(i,\text{broad})}
\end{equation}
with
\begin{align*}
\left \langle \bar{n} \right \rangle^\m{(i,broad)}&=\left(2-\left \langle
\bar{n} \right
\rangle^\m{(i,0,0)}\right)\frac{\Gamma}{2\pi}\frac{d}{d\e}\mathrm{Re}\left[\Psi\left(\frac{1}{2}+\frac{i\beta}{2\pi}\e\right)\right]\\
&\quad+\left\langle \bar{n} \right
\rangle^\m{(i,0,0)}\frac{\Gamma}{2\pi}\frac{d}{d\e}\mathrm{Re}\left[\Psi\left(\frac{1}{2}+\frac{i\beta}{2\pi}(\e+U)\right)\right]\, ,
\end{align*}
where $\Psi$ is the digamma function. Unlike the lowest non-vanishing order the first flux-dependent correction of the charge is even around $\bar{\e}=-U/2$ (see Fig \ref{reccor}).


In the linear-response limit, the transferred charge due to rectification is even with respect to magnetic flux $\phi$.
On the contrary, the flux-dependent part of the pumped current is odd. 
The different symmetry with magnetic field is related to the different processes which contribute to pumping and to rectification. For rectification, the processes that contribute to transport describe charge transfer from one lead to the other. The flux-dependent parts are associated with the interference of cotunneling through the dot and direct tunneling through the reference arm.
This is different for pumping, as discussed in the previous section. 
There, the relevant processes change the dot occupation and are resonant with intermediate state on one of the leads.
They are associated to interference between direct tunneling from the dot to a given lead and cotunneling from the dot to this lead via the other lead. 

Furthermore, the symmetry with respect to $\bar{\e}$ is different for pumping and rectification.
The pumped charge is even about $\bar{\e}= -U/2$, while for rectification there is a different symmetry in different orders of the perturbation expansion.

%double-dot ab interferometer
%adiabatic pumping
\subsection{Double-dot AB interferometer}
\label{results2dots}
For the double-dot interferometer, we consider two different limits regarding the Coulomb-interaction strength: (i) fully non-interacting case and (ii) infinite intra-dot interaction, which forbids double occupation of a single dot, and negligible inter-dot interaction.

\subsubsection{Weak adiabatic pumping}

We consider the tunneling barriers between dot and lead to be the same for both dots: $\Gamma_{\text{L}1}=\Gamma_{\text{L}2}=\Gamma_\m{L}$ and $\Gamma_{\text{R}1}=\Gamma_{\text{R}2}=\Gamma_\m{R}$. We assume the difference $\Delta \e =\e_1-\e_2$ between the dot level of the upper and of the lower dots to be of the same order as $\Gamma$.
The average level is defined as $\e = (\e_1+\e_2)/2$.
We calculate the current in first order in $\Gamma$. This order of perturbation theory is already  flux dependent. 
In the non interacting case, we can consider spinless electrons and take into account spin degeneracy by multiplying the current by a factor of 2. The dot Hilbert space for spinless electrons is spanned by 
the states $\{|0\rangle,|1\rangle,|2\rangle, |12\rangle\} $, corresponding, respectively, to both dots being empty, only upper dot occupied, only lower dot occupied, and both dots occupied. 
On the other hand, for infinite intra-dot interaction the dot Hilbert space has dimension 9 and it is spanned by the states 
$\{|0\rangle, |j\sigma\rangle, |1\sigma2\sigma'\rangle\}$, with $j=1,2$ and $\sigma,\sigma'=\uparrow,\downarrow$. These states correspond, respectively, to both dots being empty,  dot $j$ occupied with spin $\sigma$  
and both dots occupied  with spin $\sigma$ in dot $1$ and spin $\sigma'$ in dot 2. 
For the interacting system, we introduce the abbreviation $p^{j'}_j\equiv p^{j'\sigma}_{j\sigma}.$ 
In both cases addressed here, the pumped current is written conveniently as a function of the 
isospin's expectation value
\begin{align*}
\mathbf{I}=\left(\begin{array}{c}
I_x\\
I_y\\
I_z
\end{array}\right)=\frac{1}{2}\left(\begin{array}{c}
p^1_2+p^2_1\\
\m{i}(p^2_1-p^1_2)\\
p_2^2-p_1^2
\end{array}\right).
\end{align*}
Notice that if the isospin lies in the $xy$ plane, it indicates that the system is in a superposition of two states: one with only an electron in the first dot and a second  with only an electron in the second dot (both states with the same spin).  

Computing the adiabatic correction to the reduced density matrix and inserting it in Eq.~(\ref{adiabcurr}) in isospin notation, we obtain for noninteracting electrons
\begin{align}
J_{\m{L}(U=0)}^\m{(a,0)}=&-4{e}\frac{\Gamma_\m{L}}{\Gamma}\frac{2\Gamma_\m{R}\sin\frac{\phi}{2}\left(\Gamma\sin\frac{\phi}{2}+\Delta\e\cos\frac{\phi}{2}\right)+\Delta\e^2}{4\Gamma_\m{L}\Gamma_\m{R}\sin^2\frac{\phi}{2}+\Delta\e^2}\nonumber\\
&\times\frac{d f(\e)}{d\e}\frac{d\e}{d t}\, .
\end{align}
In the case of infinite interaction within one dot and vanishing interaction between the dots the expression for the current is quite long. For symmetric tunnel-coupling strengths ($\Gamma_\m{L}=\Gamma_\m{R}$) it simplifies to
\begin{align}
&J_{\m{L}(U\rightarrow\infty)}^\m{(a,0)}=-2{e}\frac{d f(\e)}{d\e}\frac{d\e}{d t}
\nonumber\\
&\times\frac{\Gamma^2\sin^2\frac{\phi}{2}\left[1+f(\e)\right]^3+\Delta\e^2[1+f(\e)]+\Gamma\Delta\e\cos\frac{\phi}{2}\sin\frac{\phi}{2}}{[1+f(\e)]^3\left\{(\Gamma^2\sin\frac{\phi}{2}[1+f(\e)]^2+\Delta\e^2\right\} }\, .
\end{align}

Now we consider weak pumping with the parameters $\e(t)=\bar\e+\delta\e(t)$ and $\Delta \epsilon(t)=\overline{\Delta\e}+\delta\Delta\e(t)$. 
The area of the cycle is  
\begin{align*}
\eta_{\Delta\e,\e}=\int \limits_0^{2\pi/\Omega} \frac{\partial
\delta\e}{\partial t}\delta \Delta\e d t.
\end{align*}
For noninteracting electrons  the pumped charge per period reads
\begin{align}
Q^\m{(a,0)}_{\Delta\e,\e(U=0)}&=8{e}\eta_{\Delta\e,\e}\frac{\Gamma_\m{L}\Gamma_\m{R}}{\Gamma}\sin\frac{\phi}{2}\frac{df(\bar{\e})}{d\bar{\e}}\nonumber \\
\times&\frac{\cos\frac{\phi}{2}(\left.\overline{\Delta\e}\right.^2-4\Gamma_\m{L}\Gamma_\m{R}\sin^2\frac{\phi}{2})-2(\Gamma_\m{L}-\Gamma_\m{R})\sin\frac{\phi}{2}\left.\overline{\Delta\e}\right.}{\left(4\Gamma_\m{L}\Gamma_\m{R}\sin^2\frac{\phi}{2}+\left.\overline{\Delta\e}\right.^2\right)^2}\, .
\end{align}
For an infinite Coulomb interaction we give, again, the expression for symmetric tunnel coupling,
\begin{align}
\label{pumpinf}
Q^\m{(a,0)}_{\Delta\e,\e(U\rightarrow\infty)}&=2{e}\eta_{\Delta\e,\e}\Gamma\frac{df(\bar{\e})}{d\bar{\e}}\nonumber \\
&\times\frac{\cos\frac{\phi}{2}\sin\frac{\phi}{2}\{\left.\overline{\Delta\e}\right.^2-\Gamma^2\sin^2\frac{\phi}{2}\left[1+f(\bar{\e})\right]^2\} }{\left[1+f(\e)\right]^3\left\{\Gamma^2\sin^2\frac{\phi}{2}\left[1+f(\e)\right]^2+\left.\overline{\Delta\e}\right.^2\right\}^2}\, .
\end{align}

\begin{figure}
\centering
\includegraphics[width=8cm]{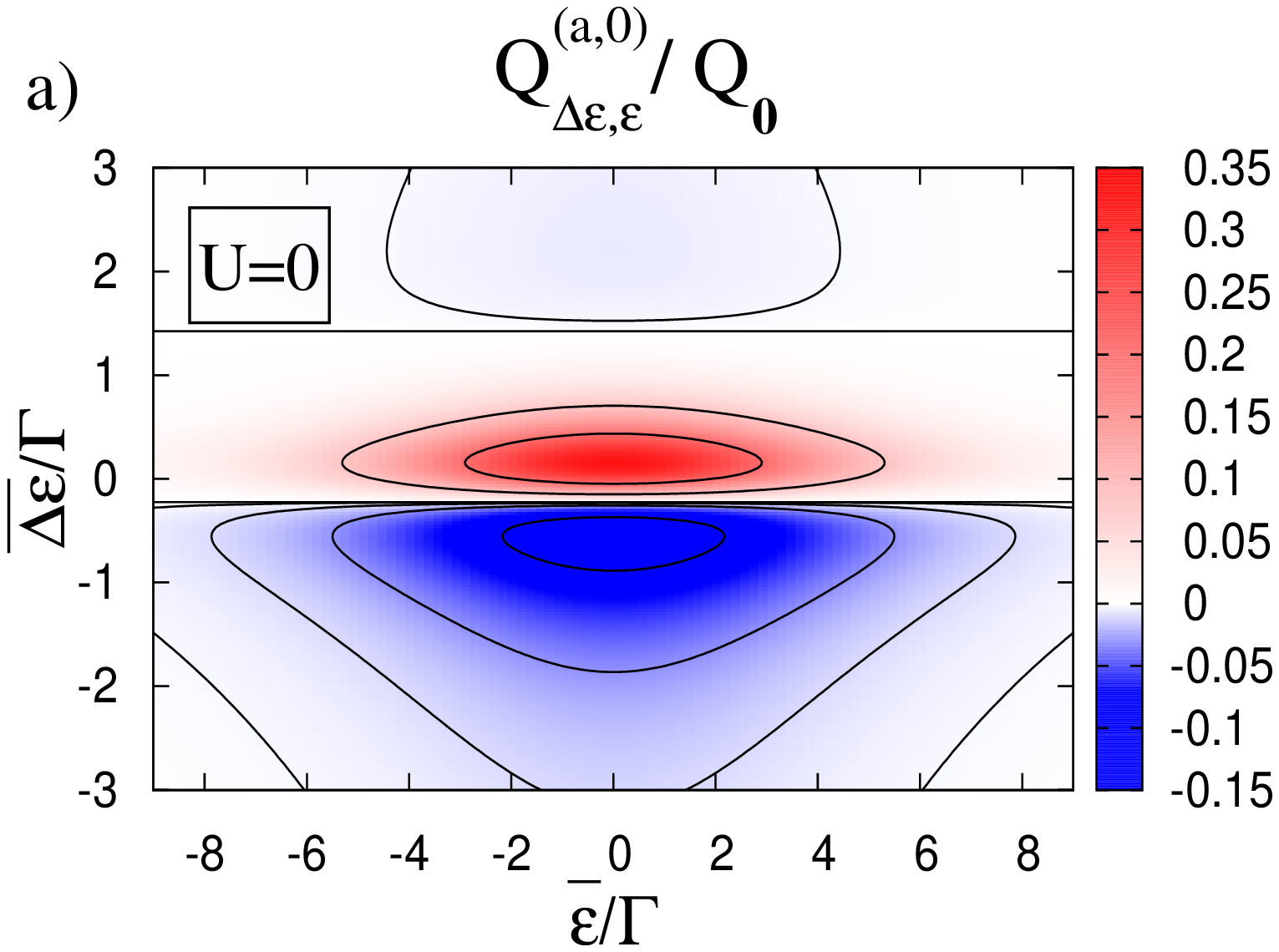}
\includegraphics[width=8cm]{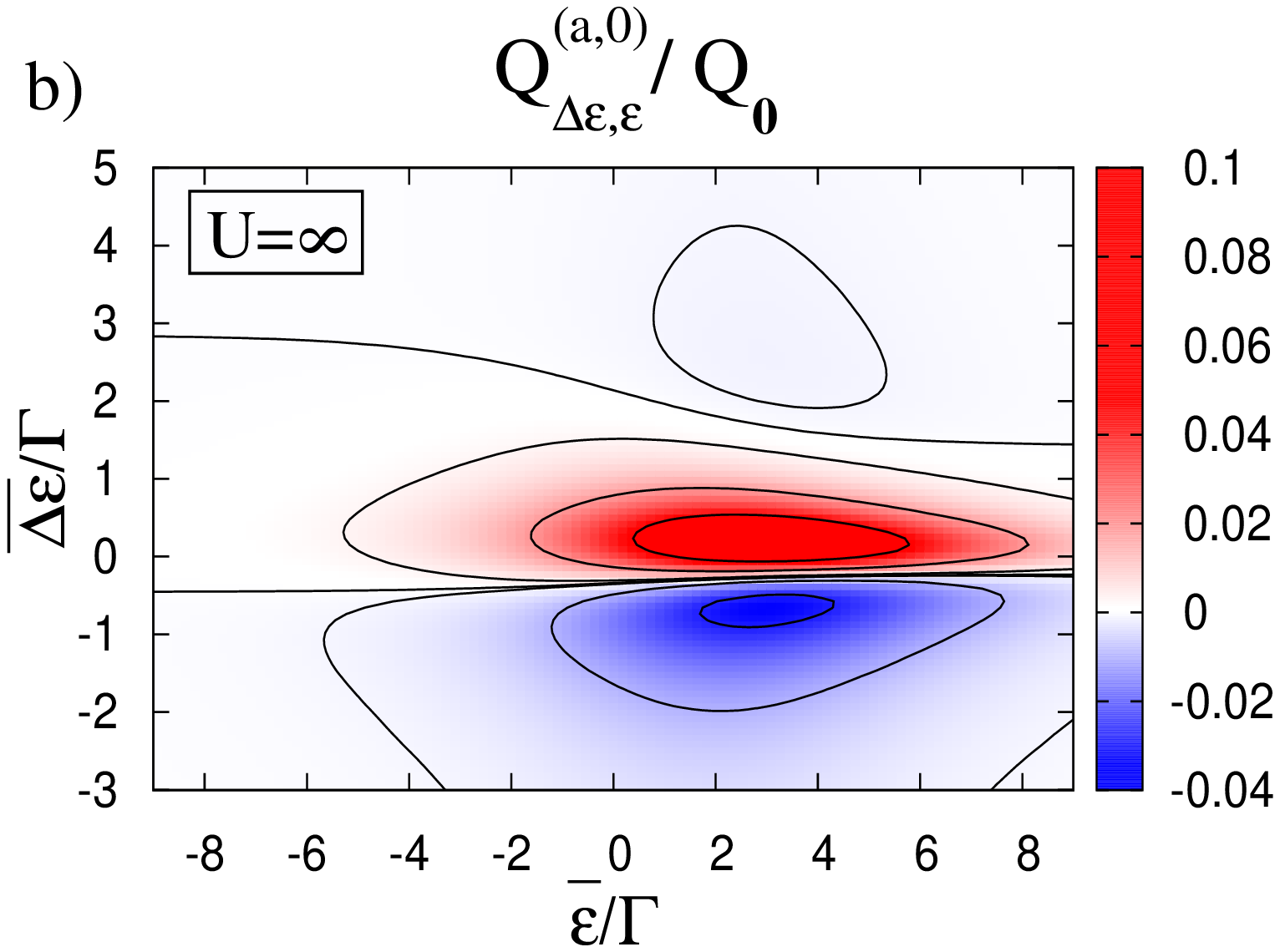}
\includegraphics[width=7cm]{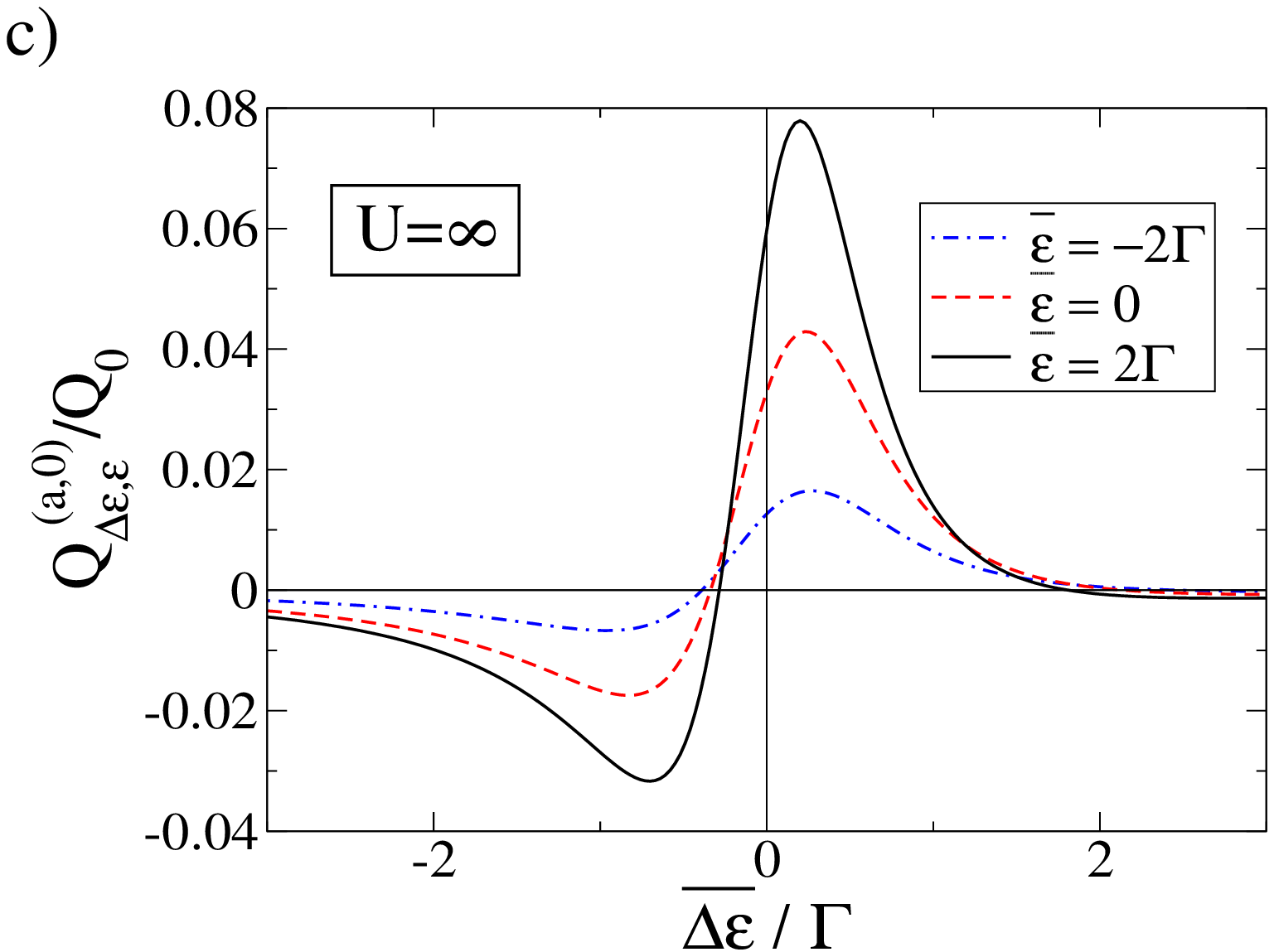}
\includegraphics[width=7cm]{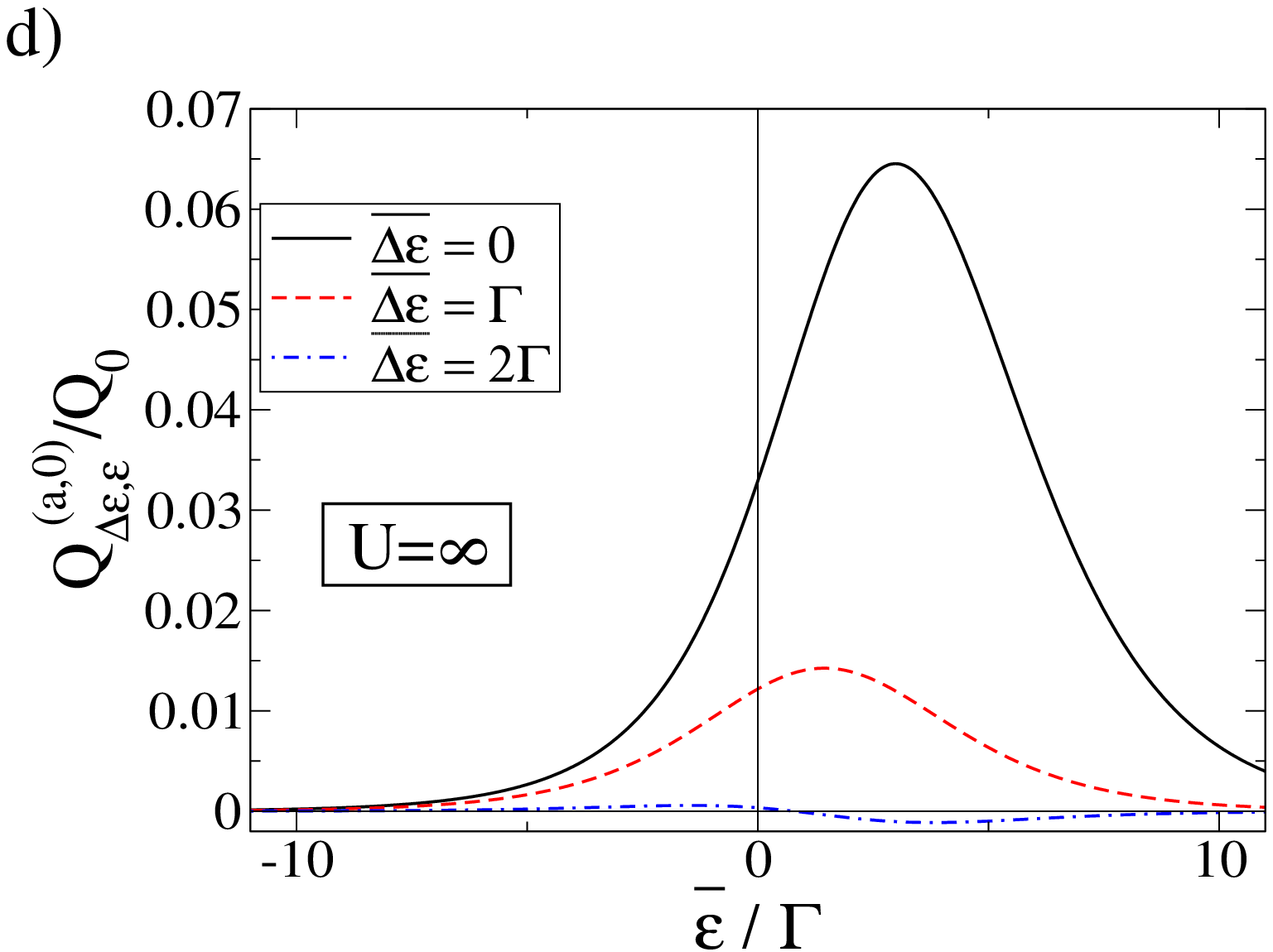}
\caption{\label{densityplot}Pumped charge
$Q^\m{(a,0)}_{\Delta\e,\e}$ in units of
$Q_0=e\eta(\e,\delta\e)/\Gamma^2$ for $\Gamma_\m{L}=0.8\Gamma$, $\Gamma_\m{R}=0.2\Gamma$, $\phi=\pi/2$, and  $k_\m{B}T=2\Gamma$. (a) and (b) show a density plot where $Q^\m{(a,0)}_{\Delta\e,\e}$ is a function of  $\bar{\e}$ and $\overline{\Delta\e}$ for (a) vanishing and (b) infinite Coulomb interaction. Cuts through (b) are shown in (c) and (d). In (c) $Q^\m{(a,0)}_{\Delta\e,\e}$ is plotted as a function of $\overline{\Delta\e}$ for different $\bar{\e}$. In (d) $Q^\m{(a,0)}_{\Delta\e,\e}$ is plotted as a function of $\bar{\e}$ for different $\overline{\Delta\e}$.
} 
\end{figure}

The pumped charge for $\Gamma_\m{L}\neq \Gamma_\m{R}$ as a function of $\bar\e$ and $\overline{\Delta\e}$ is shown in Figs.~\ref{densityplot}(a) and 5(b). We find a sign change in the pumped charge which in the noninteracting case only depends  on $\overline{\Delta\e}$ but for an infinite interaction also depends on $\overline{\e}$ . Equation \ref{pumpinf} suggests an even symmetry concerning $\overline{\Delta\e}$. Figure \ref{densityplot}c) shows that this symmetry is not general but only valid for symmetric tunneling barriers. As a function of $\bar{\e}$, the pumped charge is even only in the noninteracting but not in the interacting case, Fig.~\ref{densityplot}(d).

The fact that we find a nonvanishing pumped charge at all is not self-evident. The two pumping parameters are associated with the different arms of the interferometer. This suggests that pumping relies on coherent superposition of states localized in the different arms described by the isospin components $I_\m{x}$ and $I_\m{y}$. Therefore, one can view pumping in this case as fully quantum mechanical.

%rectification
\subsubsection{Comparison with rectification}
Similarly to the case of a single-dot interferometer, we consider rectification in the linear-response regime, in which the linear conductance, and, therefore, also the transferred charge, is an even function of the magnetic flux.
The linear conductance for vanishing interaction reads 
\begin{eqnarray}
G_{\m{L}(U=0)}^\m{(1)}=-4{e}^2\frac{\Gamma_\m{L}\Gamma_\m{R}}{\Gamma}\frac{{d}f}{{d}\e}\frac{\Delta\e^2+\Gamma_\m{L}\Gamma_\m{R}\sin^2\phi}{\Delta\e^2+\Gamma_\m{L}\Gamma_\m{R}\sin^2\frac{\phi}{2}}\, ,
\end{eqnarray}
while for infinite intra-dot interaction it is 
\begin{align}
G&_{\m{L}(U\rightarrow\infty)}^\m{(1)}=-4{e}^2\frac{\Gamma_\m{L}\Gamma_\m{R}}{\Gamma}\frac{1}{1+f(\e)} \frac{{d}f}{{d}\e}\nonumber\\
\times& \frac{\Gamma_\m{L}\Gamma_\m{R}(1-\cos\phi)\left[1+\cos\phi+2f(\e)\left(2+f(\e)\right)\right]+\Delta\e^2}{2\Gamma_\m{L}\Gamma_\m{R}(1-\cos\phi)\left(1+f(\e)\right)^2+\Delta\e^2}\, .
\end{align}

\begin{figure}
\centering
\includegraphics[width=6cm]{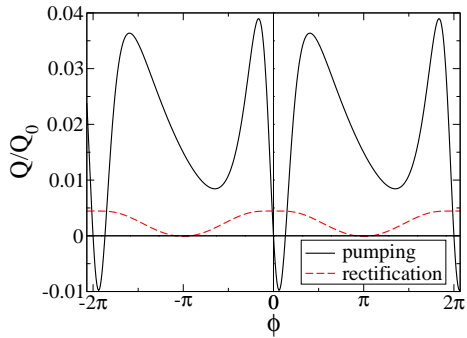}
\caption{\label{pumpfluss}Pumped charge $Q^\m{(a,0)}_{\Delta\e,\e}$ in units of
$Q_0=e\eta(\e,\delta\e)/\Gamma^2$ and rectified charge $Q^{(i,1)}_{\m{rec},\e}$ in units of $Q_0=e^2\eta_\m{rec,\e}/\Gamma$ as a function of $\phi$  for $U=\infty$, $\Gamma_\m{L}=0.8\Gamma$, $\Gamma_\m{R}=0.2\Gamma$, $\bar{\e}=0$, $\overline{\Delta\e}=0.5\Gamma$, and  $k_\m{B}T=2\Gamma$.}
\end{figure}

The pumping, on the other hand, has no definite symmetry with respect to magnetic field (Fig.~\ref{pumpfluss}) unless a symmetric choice of the tunnel-coupling strengths is assumed. 
Furthermore, we remark that the pumped charge vanishes for zero flux.

%conclusions
\section{Conclusions}\label{Conclusions}
We have investigated adiabatic pumping through an AB interferometer with a quantum dot embedded either in  one or in both arms, by means of a diagrammatic real-time approach to pumping. 
In the single-dot AB interferometer, we have found that adiabatic pumping has a peristaltic character. Nonetheless, it is clearly phase coherent as indicated by the flux dependence of the pumped current.  
On the other hand, in a double-dot AB-interferometer adiabatic pumping with the levels of the two dots is a pure quantum-mechanical transport mechanism, since it relies on the system being in a coherent superposition of eigenstates of the dots in the upper and lower arms. 
This pumping mechanism has no classical counterpart.
Finally, we found that the symmetry of the pumped charge with respect to the magnetic flux may help to distinguish pumping from rectification, at least in the linear-response regime. 
\section{Acknowledgements.}
We acknowledge financial support from the EU under Grant No. 238345 (GEOMDISS) and the DFG-Schwerpunktprogramm 1285.


\begin{thebibliography}{24}
\bibitem{note} 
The matrix elements of $\mathbf{W}(t,t')$ and $\mathbf{E}(t)$ are arranged in such a way that the matrix indices, characterized by two dot-state labels $\chi_1$ and $\chi_2$ each, match the order chosen for the vector $\boldsymbol{\pi}$.
This ensures that $W_{\chi'\chi'''}^{\chi\,\,\,\chi''}(t,t')$ and $E_{\chi'\chi'''}^{\chi\,\,\,\chi''}(t)$ connect the initial $p_{\chi'''}^{\chi''}$ with the final $p_{\chi'}^{\chi}$ state.
\end{thebibliography}
\end{document}